\documentclass[twocolumn,superscriptaddress,showpacs,floatfix,aps,prb]{revtex4}
\usepackage{graphicx}
\usepackage{dcolumn}
\usepackage{amsmath}
\usepackage{amssymb}
\usepackage{amsbsy}
\usepackage{textcomp}
\usepackage{units}
\usepackage{color}
\usepackage{units}
\usepackage[normalem]{ulem}
\usepackage{verbatim}
\usepackage{hyperref}
\usepackage{url}

\usepackage{multirow}

\usepackage{float}

\newcommand{\CaK}{CaKFe$_4$As$_4$}

\begin{document}

\title{Trends in pressure-induced layer-selective half-collapsed tetragonal
phases in the iron-based superconductor family $Ae{}A$Fe$_4$As$_4$}

\author{Vladislav Borisov}
\affiliation{Institute of Theoretical Physics, Goethe University Frankfurt am Main, D-60438 Frankfurt am Main, Germany}
\email[Corresponding author:\ ]{borisov@itp.uni-frankfurt.de}

\author{Paul C. Canfield}
\affiliation{Ames Laboratory US DOE, Iowa State University, Ames, Iowa 50011, USA}
\affiliation{Department of Physics and Astronomy, Iowa State University, Ames, Iowa 50011, USA}

\author{Roser Valent\'i}
\affiliation{Institute of Theoretical Physics, Goethe University Frankfurt am Main, D-60438 Frankfurt am Main, Germany}

\date{\today}

\begin{abstract}

By performing pressure simulations within density functional theory for the
family of iron-based superconductors $Ae{}A$Fe$_4$As$_4$ with $Ae$ = Ca, Sr, Ba
and $A$ = K, Rb, Cs we predict in these systems the appearance of two consecutive 
half-collapsed tetragonal transitions at pressures $P_{c_1}$ and $P_{c_2}$, which 
have a different character in terms of their effect on the electronic structure. 
We find that, similarly to  previous studies for \CaK{}, spin-vortex magnetic fluctuations 
on the Fe sublattice play a key role for an accurate structure prediction in these materials 
at zero pressure. We identify clear trends of critical pressures and discuss 
the relevance of the collapsed phases in connection to magnetism and superconductivity. 
Finally, the intriguing cases of EuRbFe$_4$As$_4$ and EuCsFe$_4$As$_4$, where Eu magnetism 
coexists with superconductivity, are discussed as well in the context of half-collapsed phases.

\end{abstract}

%\pacs{}

\maketitle

%\section{Introduction}

\begin{figure}
  \includegraphics[width = 0.99\columnwidth]{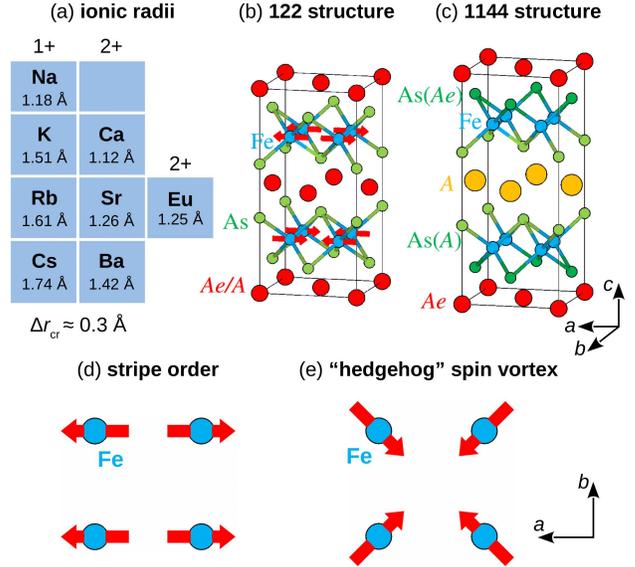}
  \vspace{-20pt}
  \caption{
  (a) Cations of alkali ($1+$) and alkaline-earth ($2+$) elements, as well as 
  divalent Eu, together with their ionic radii from Ref.~\onlinecite{Iyo2016,Shannon1976}. 
  (b) The 122 and (c) 1144 structures of iron pnictides. In general, 
  the 1144 phase of $Ae{}A$Fe$_4$As$_4$ is stable when the difference in the ionic 
  radii $\Delta r = |r(Ae) - r(A)|$ is larger than 0.3\,\AA. Possible Fe magnetic orders 
  are shown in (d) stripe order and (e) ``hedgehog'' or spin-vortex order.}
  \label{f:122_vs_1144}
  \vspace{-15pt}
\end{figure}

\textit{Introduction.--} 
The so-called 122 Fe-based pnictides (Fig.~\ref{f:122_vs_1144}b) ($A$Fe$_2$As$_2$, 
$Ae$Fe$_2$As$_2$, EuFe$_2$As$_2$) with \textit{A} alkali and \textit{Ae} alkaline-earth 
cations crystallize at room temperature in a body-centered tetragonal ThCr$_2$Si$_2$ 
structure ($I4/mmm$)~\cite{Kreyssig2008,Canfield2010} where As sites from the neighboring 
Fe-As blocks face each other across the \textit{A} or \textit{Ae} plane. The As-As interlayer 
distance in these systems can be then tuned by either mechanical or chemical pressure down 
to sufficiently small values allowing the formation of As-As $p_z$ bonds. This is accompanied 
by a structural phase transition to a collapsed tetra\-gonal (cT) phase where the $c/a$ ratio 
is significantly reduced due to a dramatic contraction of the $c$-lattice parameter and a slight 
expansion of the $a$-lattice parameter. This process is known to suppress superconductivity 
or/and long-range ``stripe'' magnetic order (Fig.~\ref{f:122_vs_1144}d) due to the crossover 
to a more three-dimensional structure and the loss of spin fluctuations and local Fe 
moments caused by a compression of Fe-As bonds.\cite{Yildirim2009} In the 122 materials, 
the transition to a cT phase affects the whole structure leading to  As-As $p_z$ bond 
formation across each cation spacer layer.

In contrast, a \textit{half-collapsed} tetragonal (hcT) phase transition was recently 
reported for the 1144 material \CaK{}\cite{Kaluarachchi2017} (Fig.~\ref{f:122_vs_1144}c) 
where the periodic arrangement of Ca and K spacer layers produces two different kinds of 
As sites\cite{Iyo2016,Meier2016,Budko2017} and the tetragonal structure ($P4/mmm$) shows 
a layer-selective collapse upon application of pressure. First, at 4~GPa the As-As $p_z$ 
bonding across the Ca layer induces a collapsed tetragonal transition with disappearance 
of superconductivity while a second collapsed transition across the K layer was predicted 
around $\sim$12~GPa. Furthermore, Ref.~\onlinecite{Kaluarachchi2017} showed that ``hedgehog'' 
(spin-vortex, Fig.~\ref{f:122_vs_1144}e) magnetism had to be invoked in the pressure-dependent 
density functional theory (DFT) simulations in order to predict the observed structural 
transitions. This magnetic order has been recently measured upon electron-doping 
\CaK{}.\cite{Meier2018}

The rare-earth-based EuRbFe$_4$As$_4$ and EuCsFe$_4$As$_4$ are also attracting growing 
attention due to the coexistence of Eu magnetism and bulk superconductivity as reported 
in Refs.~\onlinecite{Kawashima2016,Liu2016,Liu2017,Jackson2018}. However, the exact nature 
of the Eu magnetic order and its effect on the superconductivity remain to be elucidated, 
as well as possible Eu$^{2+}$ to Eu$^{3+}$ transitions previously observed for 122 
systems.\cite{Zapf2017} The interactions between the localized Eu spins are expected to 
be sensitive to the lattice parameters and, in this respect, the half-collapse transition 
might influence the ground state which stimulates a detailed study.

Predicting the appearance of possible cT transitions in ThCr$_2$Si$_2$-structured intermetallic 
compounds is not only of relevance for the superconducting and magnetic properties of these 
materials but also for their potential superelastic behavior as has been recently shown.\cite{Sypek2017} 

In this work we systematically study via DFT calculations possible pressure-induced 
half-collapsed tetragonal transitions in a series of previously synthesized 1144 
systems\cite{Iyo2016} with the following combinations of the spacer cations: 
CaRb, CaCs, SrRb, SrCs, and BaCs,\footnote{We note that the exact crystal structure 
of BaCsFe$_4$As$_4$ is not completely clarified and this material can probably exist 
both in the 122 and in the 1144 forms due to a smaller difference in the cation radii, 
compared to other 1144 systems. Nevertheless, we include the 1144 phase of BaCsFe$_4$As$_4$ 
in our simulations, in order to reveal general trends in the 1144 family of iron pnictides.} 
as well as EuRb and EuCs. We discuss the tendencies expected for the transition pressures 
in relation to the nature of the spacer cations, the underlying Fe-moment fluctuations and 
possible magnetism in Eu for the latter two systems.

\begin{figure}[H]
  \includegraphics[width = 0.99\columnwidth]{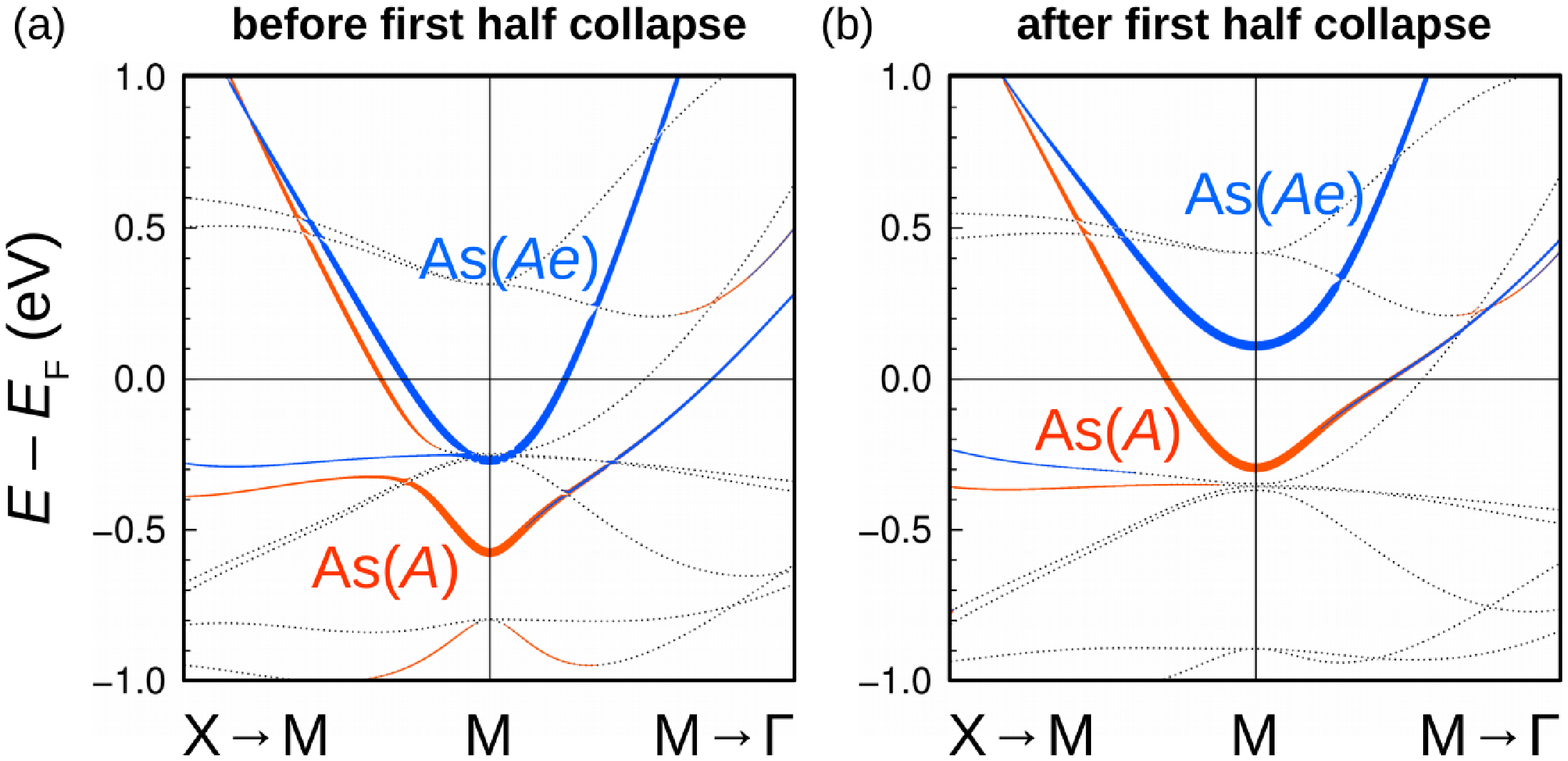}
  \includegraphics[width = 0.99\columnwidth]{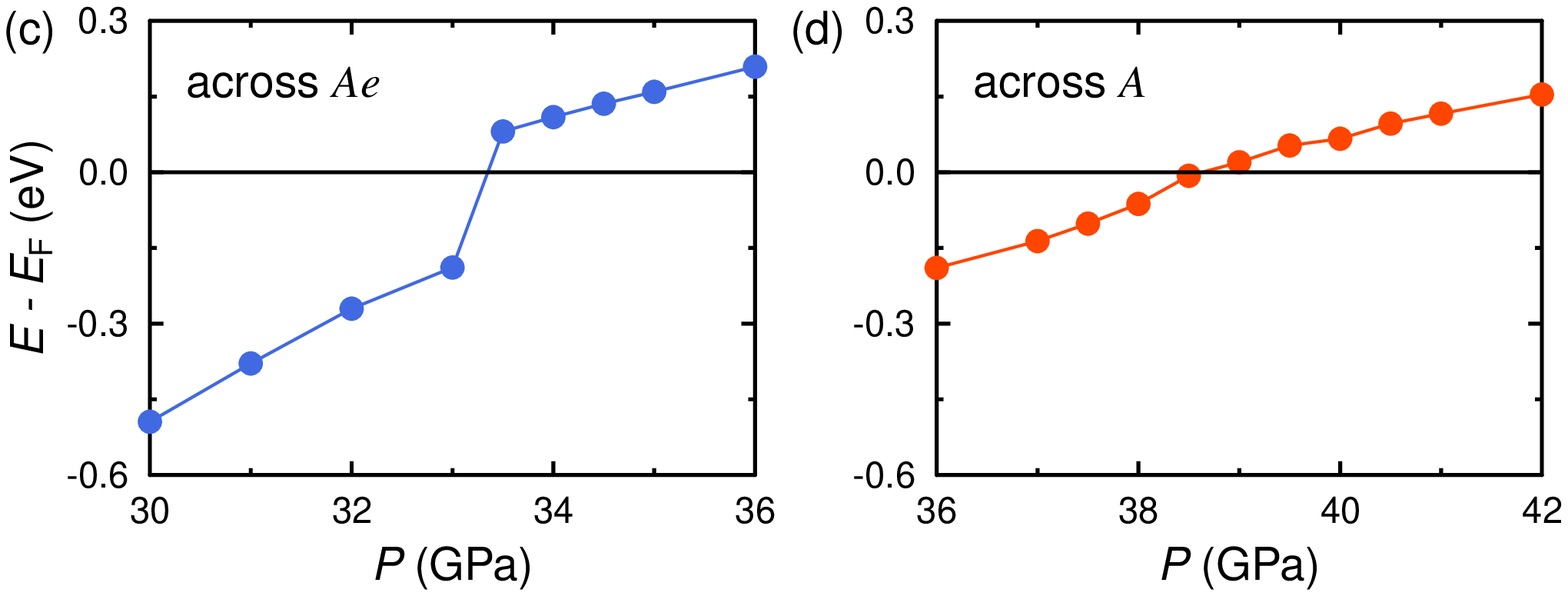}
  \vspace{-25pt}
  \caption{
  (a,b) Characteristic evolution of the non-spin-polari\-zed band structure and 
  (c,d) the energy position $E$ of the antibonding As-As $p_z$ orbital relative to the 
  Fermi energy $E_\mathrm{F}$ across a hcT transition in \textit{Ae}\textit{A}Fe$_4$As$_4$ 
  (BaCsFe$_4$As$_4$ data are taken here as an example). Upon the first hcT, only the 
  As-As $p_z$ antibonding band from	the As facing the smaller cation layer, in this 
  case \textit{Ae} (blue lines), shifts abruptly above the Fermi level (plot c), 
  while the band from As facing the larger cation layer (orange lines) remains occupied. 
  This suggests that As-As bonds are strongly formed across the \textit{Ae} layer. 
  The second hcT at higher pressures is identified in the same fashion and, in contrast 
  to the first hcT, reveals a smooth shift of the corresponding As bands across the 
  Fermi level (plot d).
  }  \label{f:collapse_criterium}
\end{figure}

\vspace{10pt}
\textit{Methods.--} Structural transitions under pressure are simulated using the 
projector-augmented wave method\cite{Bloechl1994,Kresse1999} as implemented 
in the VASP code~\cite{Kresse1993,Kresse1996,Kresse1996b} within the GGA 
exchange-correlation approximation.\cite{Perdew1996} For the two systems with Eu, 
strong correlations for the \textit{4f} states were included using two different 
GGA+$U$ schemes.\cite{Dudarev1998,Liechtenstein1995} Convergence of the properties 
of interest is achieved for a $(5\times 5\times 5)$ $k$-mesh and an energy cutoff 
of 600~eV. Increase of the cutoff up to 800~eV changes the $c$-lattice parameter 
by $\sim$0.05\,\AA{} and the critical pressure by less than 0.5~GPa (see 
Fig.~\ref{f:energy_cutoff_effect} in Appendix B), which is an acceptable accuracy 
for studying the pressure-structure trends in the 1144 systems.

At low pressures, the structural prediction is done while imposing a
``frozen'' spin-vortex configuration of Fe moments (Fig.~\ref{f:122_vs_1144}e),
which simulates, to a first approximation, the effect of spin fluctuations, as
shown in our previous work on \CaK{}.\cite{Kaluarachchi2017,Cui2017} The assumption of a
particular underlying Fe magnetism is necessary for a correct prediction of the 
collapsed tetragonal transition, even if the actual system doesn't show any 
long-range magnetic order. Purely non-magnetic calculations in 
Fe-based superconductors fail to reproduce the correct structural 
parameters.\cite{Zhang2009,Colonna2011,Tomic2012,Dhaka2014,Budko2016} 

Once the optimized 1144 structures under pressure are obtained, the electronic 
bands are calculated using the all-electron full-potential localized orbitals 
basis set (FPLO) code\cite{FPLO} with the GGA exchange-correlation functional
(GGA+$U$ for Eu-based systems). The critical pressure for a hcT transition is captured 
by monitoring the energy position of the antibonding As-As 4$p_z$-based molecular 
orbitals, which shift towards the Fermi level with increasing pressure. At the 
transition, these bands shift abruptly above the Fermi level (compare
Fig.~\ref{f:collapse_criterium} (a) and (b)) and the corresponding As-As $p_z$ bonds 
are enhanced. This criterium was successfully applied to several \textit{Ae}Fe$_2$As$_2$ 
systems (\textit{Ae}=Ca,\,Sr,\,Ba),\cite{Yildirim2009,Colonna2011,Kasinathan2011,Tomic2012,Dhaka2014,Diehl2014}
as well as to \CaK{}.\cite{Kaluarachchi2017} As we will discuss below, the As band 
shift corresponding to the second hcT phase at higher pressures is smoother than 
the one corresponding to the first hcT phase as shown in Fig.~\ref{f:collapse_criterium} 
(c) and (d).

\begin{figure}
  \includegraphics[width = 0.95\columnwidth]{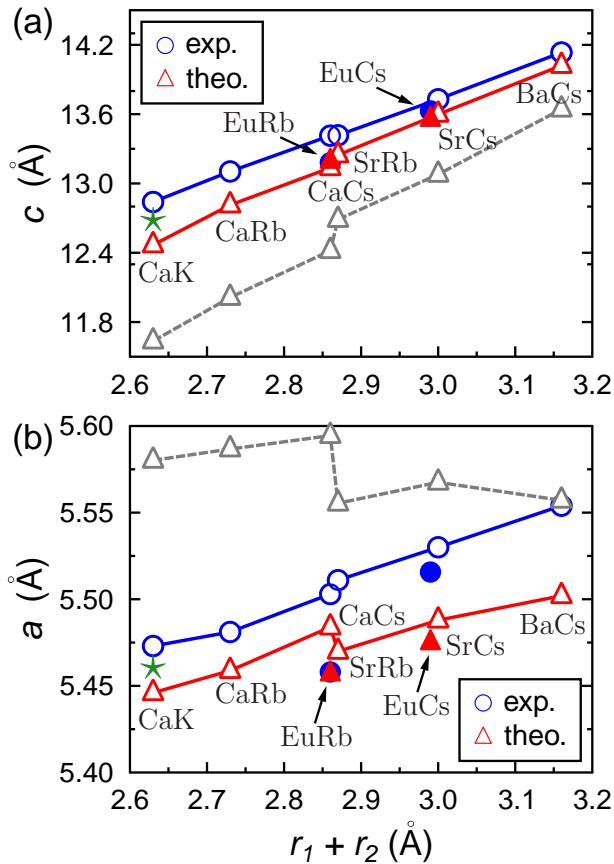}\vspace{-15pt}
  \caption{Correlation between the zero-pressure $a$- and $c$-lattice parameters
and the sum of the ionic radii $r_1 + r_2$ of the spacer cations ($r_1$ for
\textit{Ae} and Eu and $r_2$ for \textit{A} cations) for different 1144 iron
pnictides. The measured (circles) and theoretical values (triangles) are shown
for each system. Filled symbols indicate the parameters of the Eu-based systems
and the star shows the low-temperature data for \CaK{}.\cite{Meier2016} The theory 
prediction is obtained using GGA and the spin-vortex configuration of Fe moments,
following previous work,\cite{Kaluarachchi2017} from which the data for \CaK{}
is taken for this plot. Dashed lines show results of purely non-magnetic calculations.}  \label{f:c_exp_vs_theo}
\end{figure}

\textit{Results (zero pressure).--} For all systems studied here we find that 
the assumption of a spin-vortex-type magnetism in our DFT calculations is crucial 
to reproduce the experimental structure at zero pressure as shown in 
Fig.~\ref{f:c_exp_vs_theo} where the measured $c$- and $a$-lattice parameters are 
compared with the optimized ones. In contrast, $c$ is severely underestimated 
by $\sim$(0.5\,--\,1.0)\,\AA{} and $a$ is overestimated by $2\%$ in non-magnetic 
calculations (dashed lines in Fig.~\ref{f:c_exp_vs_theo}). We note, however, that 
the theoretical values\- for the $c$ parameter still deviate by (0.1\,--\,0.3)\,\AA{} 
from the experimental result, where smaller deviations are observed for compounds 
with larger spacer cations. One possible source of discrepancy is the fact that the 
1144 crystal structures were measured at room temperature,\cite{Iyo2016} while DFT 
calculations formally correspond to the zero-temperature case, suggesting that thermal 
expansion can partially explain the observed deviations. For example, the thermal 
effect\- in the $c$ parameter of \CaK{} can be as large as 0.16\,\AA{} when going 
from room temperature down to a few kelvins,\cite{Meier2016} so that the overall 
agreement between theory and experiment for this compound is acceptable 
(compare circles and triangle symbols in Fig.~\ref{f:c_exp_vs_theo}).

\begin{figure*}
  \includegraphics[width = 0.99\textwidth]{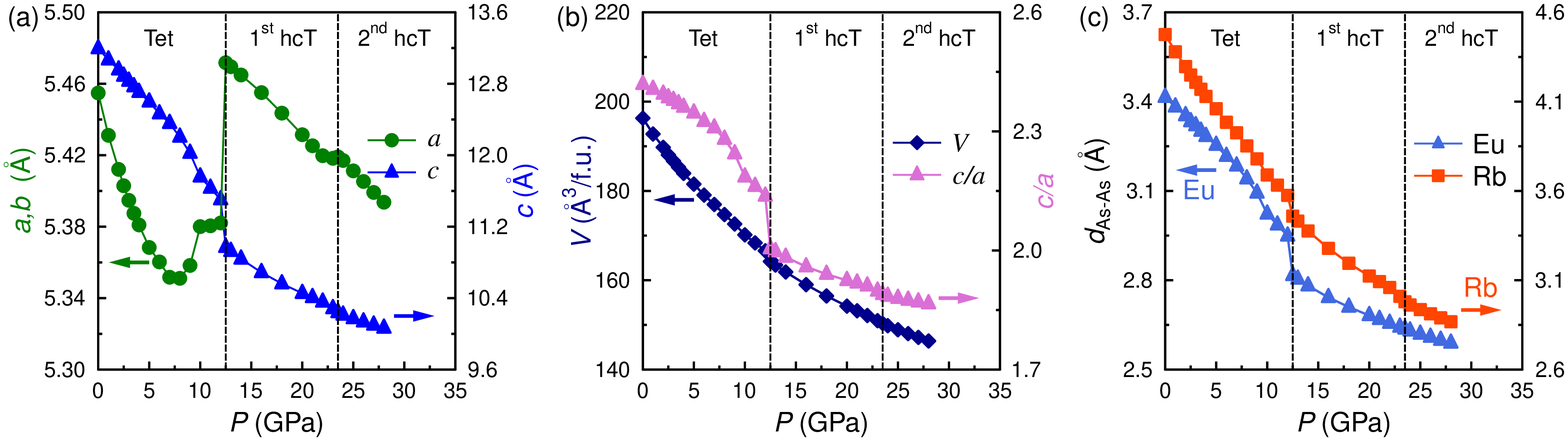}%\vspace{-5pt}
  \caption{Pressure evolution of (a) lattice parameters $a=b$ and $c$, (b) volume and $c/a$ ratio and (c) As-As distances across both hcT transitions for EuRbFe$_4$As$_4$. The critical pressures of the two half-collapsed transitions are marked by vertical dashed lines. Here, the first hcT and the collapse of Fe moments occur simultaneously.} \label{f:EuRb_structure_vs_pressure_1}
\end{figure*}

\begin{figure*}
  \includegraphics[width = 0.99\textwidth]{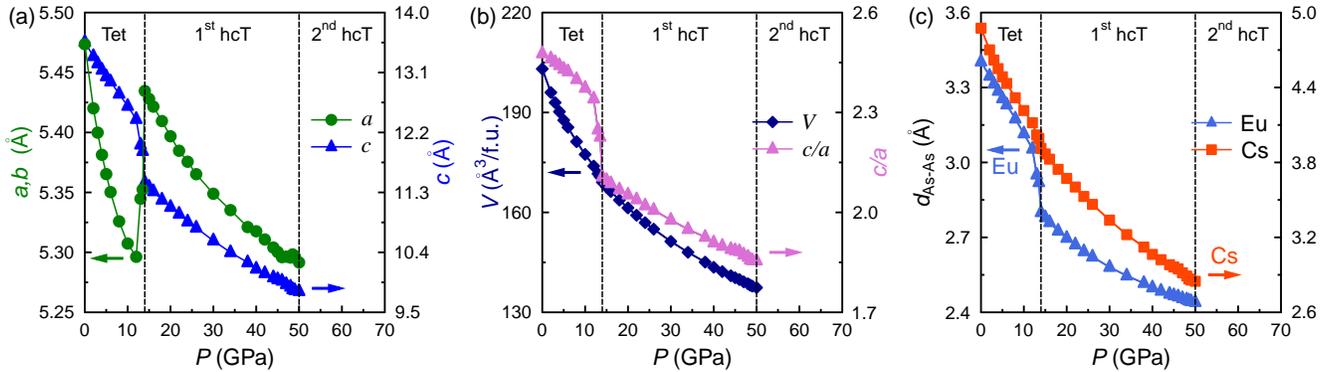}\vspace{-5pt}
  \caption{Pressure evolution of (a) lattice parameters $a=b$ and $c$, (b) volume and $c/a$ ratio and (c) As-As distances across both hcT transitions for EuCsFe$_4$As$_4$. The critical pressures of the two half-collapsed transitions are marked by vertical dashed lines. Here, the first hcT and the collapse of Fe moments occur simultaneously.} \label{f:EuCs_structure_vs_pressure}
\end{figure*}

\textit{First half-collapsed tetragonal transition.--} Assuming a spin-vortex magnetic 
configuration, we performed pressure simulations for all 1144 materials over a wide 
range of pressures. The observed evolution of the in-plane, $a$, and
out-of-plane, $c$, lattice parameters with applied pressure is qualitatively
similar in all cases, despite large variations in their absolute values 
(see Figs.~\ref{f:EuRb_structure_vs_pressure_1},~\ref{f:EuCs_structure_vs_pressure} 
as well as Figs.~\ref{f:CaRb_structure_vs_pressure}-\ref{f:BaCs_structure_vs_pressure} 
in Appendix C). At the first  hcT transition, the $c$ parameter and the As-As interlayer 
distances ($d_\mathrm{As-As}$) across the cation plane where the $c$ collapse is happening 
decrease abruptly, while the in-plane parameter $a$ shows an upturn. The As-As $4p_z$ 
antibonding orbitals lie above the Fermi level as illustrated in Fig.~\ref{f:collapse_criterium} 
(a) and (c). As a matter of fact, the As-As distance right after the first hcT is very similar 
for all 1144 compounds and varies between (2.8\,--\,2.9)\,\AA{} (Table~I), which agrees with 
the chemical nature of collapsed tetragonal transitions in these materials.

The estimated critical pressure for the first half-collapsed tetragonal 
transition ($P_{c1}$ in Table~I) shows clear trends as a function of the 
cation sizes. For all Ca\textit{A}Fe$_4$As$_4$ (\textit{A} = K,\,Rb,\,Cs) systems, 
the first hcT is observed near 5~GPa, which agrees with the fact that the 
As-As $p_z$ bonding is happening across the Ca layers in all three systems. 
When \textit{Ae}=Ca is replaced by Sr, $P_{c1}$ is shifted to larger values, 
up to 14--15~GPa for \textit{A}=Rb and up to 18~GPa for \textit{A}=Cs. Similar 
critical pressures for the first hcT are found here for Eu-based 1144 systems, 
which can be expected based on almost identical ionic radii of Sr and Eu (Table\-~I).
The maximal critical pressure $P_{c1}$ of 34~GPa is found for BaCsFe$_4$As$_4$ with
the largest alkaline-earth interlayer cation (Ba). From Table~I, it can be observed
that for a given $Ae$ cation, increasing the size of the \textit{A} cation shifts the
first hcT phase to higher pressures. Similarly to the 122 pnictides, the half-collapsed 
transitions in the 1144 systems are more abrupt for smaller spacer cations and become 
broadened for cations with larger ionic radii.

One point to discuss is the consequences of invoking a spin-vortex Fe magnetic order 
for the structural relaxa\-tions under pressure even though these systems at zero pressure
do not manifest magnetic order associated with the Fe site. This assumption may lead 
to slightly overestimated $P_{c1}$. Whereas we  have discussed above the importance of 
introducing this ``frozen'' magnetic order, the calculations show, on top of the structural
collapse, a ``magnetic collapse'' that is absent in the real system. Such a collapse 
may happen simultaneously to the structural collapse or at slightly different pressures 
as it is the case for CaRbFe$_4$As$_4$, CaCsFe$_4$As$_4$ or BaCsFe$_4$As$_4$ (see the grey 
area in Figs.~\ref{f:CaRb_structure_vs_pressure},~\ref{f:CaCs_structure_vs_pressure},~\ref{f:BaCs_structure_vs_pressure}). 
This result may be interpreted in terms of the hcT phase being a broad transition smeared
by spin fluctuations.

For the special cases of EuRbFe$_4$As$_4$ and EuCsFe$_4$As$_4$ we find the first hcT phase 
to occur across the Eu layer at a $P_{c1}$ of about 12.5~GPa and 14~GPa respectively
(Figs.~\ref{f:EuRb_structure_vs_pressure_1},~\ref{f:EuCs_structure_vs_pressure}). 
This prediction shows a rather good agreement with the most recent experimental study of 
pressure effects in EuRbFe$_4$As$_4$ and EuCsFe$_4$As$_4$.\cite{Jackson2018} In our simulations, 
Eu ferromagnetism\footnote{Please note that in our simulations we didn't consider more complex 
cases of Eu magnetism than ferromagnetism.} survives well beyond the first hcT as has also been 
recently observed~\cite{Jackson2018}. This indicates that the localized Eu spins are little 
influenced by the As-As $p_z$ bonding. At much higher pressures, as we discuss below, the 
second collapse occurs across the Rb plane ($P_{c2}\sim 23.5\:\text{GPa}$) for EuRbFe$_4$As$_4$ 
and across the Cs plane ($P_{c2}\sim 48.5\:\text{GPa}$) for EuCsFe$_4$As$_4$. The ferromagnetic 
order of Eu persists up to the highest studied pressure slightly above the second\- collapse. 
No drastic changes of the Eu oxidation state are observed in the whole pressure range.

\begin{table*}
  \caption{Predicted critical pressures $P_{c1}$ and $P_{c2}$ and As-As
interlayer distances $d_\mathrm{As-As}$ for the two hcT transitions
in the 1144 series (data for \CaK{} is taken from Ref.~\onlinecite{Kaluarachchi2017}). 
Note that the values provide indicative trends for the real systems. The ionic radii $r_1$ 
and $r_2$ are provided for the \textit{Ae} and \textit{A} species and Eu (same data in 
Fig.~\ref{f:122_vs_1144}a). The accuracy of the provided critical pressures lies within 
0.5~GPa, as determined by the smallest pressure step and convergence of the simulations. 
Superconducting $T_c$ from Ref.~\onlinecite{Iyo2016} and \onlinecite{Kawashima2016} 
is shown for each compound together with the zero-pressure As height asymmetry 
$\eta_0 = \eta(P = 0)$ defined below by expression (\ref{e:eta}).} \vspace{5pt}
\centering
\setlength{\tabcolsep}{7pt}
\renewcommand{\arraystretch}{1.5}
   \begin{tabular}{ccc|ccc|ccc}\hline\hline
       & & & \multicolumn{3}{c|}{first hcT} & \multicolumn{3}{c}{second
 hcT} \\ \hline
      Compound  &  $T_c$ (K)  &  $\eta_0$ (\%)  &  $r_1$ (\AA)  &  $P_{c1}$ (GPa)  & $d_\mathrm{As-As}$ (\AA) &  $r_2$ (\AA)  &  $P_{c2}$ (GPa)  & $d_\mathrm{As-As}$ (\AA) \\ \hline
  	   CaKFe$_4$As$_4$   &  33.1  &  1.47  &   1.12    &   4      &   2.82  &  1.51  &  12.4  &  3.00  \\
  	  CaRbFe$_4$As$_4$   &  35.0  &  1.48  &   1.12    &   5.25   &   2.79  &  1.61  &  26    &  2.95  \\
  	  CaCsFe$_4$As$_4$   &  31.6  &  1.54  &   1.12    &   5.8    &   2.79  &  1.74  &  58    &  2.81  \\
  	  SrRbFe$_4$As$_4$   &  35.1  &  0.76  &   1.26    &   14.5   &   2.88  &  1.61  &  24    &  2.95  \\
  	  SrCsFe$_4$As$_4$   &  36.8  &  0.72  &   1.26    &   18     &   2.78  &  1.74  &  46.5  &  2.88  \\
  	  BaCsFe$_4$As$_4$   &  26.0  &  0.43  &   1.42    &   34     &   2.91  &  1.74  &  39    &  2.97  \\
  	  EuRbFe$_4$As$_4$   &  36    &  0.76  &   1.25    &   12.5   &   2.82  &  1.61  &  23.5  &  2.98  \\
  	  EuCsFe$_4$As$_4$   &  35    &  0.73  &   1.25    &   14     &   2.80  &  1.74  &   48.5   &   2.87  \\ \hline\hline
   \end{tabular}
   \label{t:critical_pressures}
\end{table*}


\textit{Second half-collapsed tetragonal phase.--}
 At higher pressures, the studied\- 1144 systems undergo a second hcT transition 
in our simulations. In our previous work on \CaK{}, we predicted the second structural collapse 
to occur around 12~GPa.\cite{Kaluarachchi2017} Here, we find that the second critical pressure 
$P_{c2}$ correlates with the size of the corresponding spacer cation (\textit{A} = K,\,Rb,\,Cs), 
similarly to the first hcT phase where $P_{c1}$ rapidly increases in the series 
Ca,\,Sr,\,Eu,\,Ba. The lowest pressure for the second hcT is expected for \CaK{}, whereas 
the highest one (58~GPa) is found for CaCsFe$_4$As$_4$. For that reason, the se\-cond hcT 
transition might be difficult to access experi\-mentally.

Interestingly, the second hcT transition has a relatively smaller effect\- 
on the lattice parameters (only a small kink is observed for $a$ and $c$) 
and is detected in our first-principles calculations again based on the 
analysis of the As-As $4p_z$ orbital bonding, 
as demonstrated in Fig.~\ref{f:collapse_criterium}. 
The characteristic As-As distance  $d_\mathrm{As-As}$ across the \textit{A} layer 
is around $(2.8-3.0)\,\mathrm{\AA}$ after the collapse (Table~I), which is on average 
slightly larger than $d_\mathrm{As-As}$ across the \textit{Ae} layer at $P_{c1}$. 
The smallest critical $d_\mathrm{As-As}$ of 2.81\,\AA{} is found for CaCsFe$_4$As$_4$, 
which correlates with the highest critical pressure $P_{c2}$ of 58~GPa. It should be 
emphasized that, for all studied 1144 systems, the Fe moments are fully suppressed 
long before the second\- hcT transition and play no role for this  transition. 
The purely chemical nature of the second half collapse becomes then even more apparent.

\begin{figure}%[H]
\vspace{-10pt}
\includegraphics[width = 0.99\columnwidth]{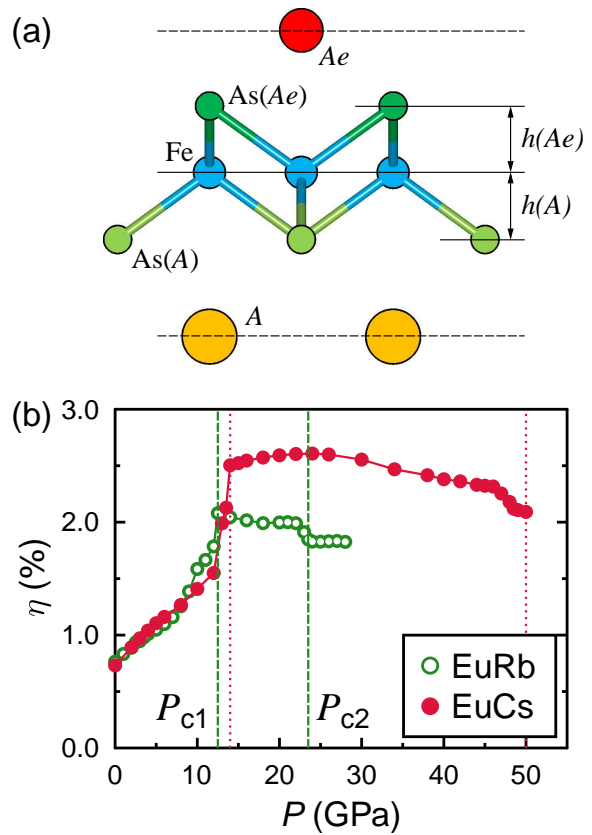}
\vspace{-10pt}
  \caption{
  (a) Definition of two different As heights in 1144 compounds. (b) As height asymmetry 
  (\ref{e:eta}) vs pressure for EuRbFe$_4$As$_4$ and EuCsFe$_4$As$_4$. 
  The critical pressures $P_{c1}$ and $P_{c2}$ for both hcT are indicated by the 
  vertical dashed (EuRb) and dotted (EuCs) lines. Similar qualitative behavior is 
  observed for other 1144 systems.
  }  \label{f:hAs_asymmetry_vs_p}
\end{figure}

\textit{As height asymmetry.--} The main difference between the well-known 122 and 
the new 1144 (\textit{AeA}Fe$_4$As$_4$) compounds is the broken glide-plane symmetry 
in the latter case, which creates two types of As sites. This asymmetry of As tetrahedra 
in the 1144 systems was found to play a leading role for the emergence of  spin-vortex 
magnetism under electron doping.\cite{Meier2018} We can characterize this structural 
property by the parameter
\begin{equation}
	\eta = \frac{ h(A) - h(Ae) }{ h(A) + h(Ae) }\times 100\%,  \label{e:eta}
\end{equation}

with $h(Ae)$ and $h(A)$ being the As heights on the side of the \textit{Ae} and 
\textit{A} spacer layers, respectively (Fig.~\ref{f:hAs_asymmetry_vs_p}a).

We find that $h(A)$ is always larger than $h(Ae)$ for all studied 1144
pnictides. Also, the largest As height asymmetry at zero pressure is found for
Ca\textit{A}Fe$_4$As$_4$ ($\eta \approx 1.5\% $), while the asymmetry gradually
decreases towards 0.4\% in the series \textit{Ae} = Ca,\,Sr,\,Ba. As evident 
from Table~I, this fact is directly related to the diffe\-rence in the ionic
radii of the \textit{Ae} and \textit{A} cations.

Upon increasing pressure, the As height asymmetry $\eta$ grows continuously
until the first hcT transition is reached where it shows a clear upturn
(Fig.~\ref{f:hAs_asymmetry_vs_p}b). The Fe magnetic collapse in the calculation 
is always accompanied by a sudden increase of the asymmetry para\-meter $\eta$, while the
second hcT transition slightly reduces the asymmetry. These trends are observed
for all studied systems.

Since a large As height asymmetry favors the ``hedgehog'' spin-vortex magnetic order in \CaK{}\cite{Meier2018} and increases under pressure for the studied 1144 systems 
(example\- in Fig.~\ref{f:hAs_asymmetry_vs_p}b), we can argue that pressure can stabilize 
the spin-vortex state relative to the usual stripe phase. On the other hand, BaCsFe$_4$As$_4$ 
is predicted here to have a more symmetric As-Fe-As block than other 1144 systems and, for 
that reason, is likely to be closer to stripe order than the other 1144 compounds.

\textit{Conclusions.--} We performed first-principles density functional theory 
simulations under pressure of various members of the 1144 family of Fe-based 
superconductors and found clear trends in the appearance of layer-selective 
half-collapsed tetragonal transitions as well as changes in the electronic properties 
of these systems. First of all, the criti\-cal pressures $P_{c1}$ and $P_{c2}$ for 
both consecutive hcT transitions increase rapidly with the cation size of the 
respective spacer layers. This agrees with the already known features of collapse 
in \textit{Ae}Fe$_2$As$_2$ systems (\textit{Ae} = Ca, Sr, Ba)\cite{Mittal2011,Kasinathan2011,
Tomic2012,Tomic2013,Guterding2017} and EuFe$_2$As$_2$.\cite{Uhoya2010,Zapf2017}.
Secondly, even though the systems considered don't show magnetic long-range order at the Fe sites, 
it is necessary to include Fe spins in a certain magnetic arrangement in simulations in order 
to correctly predict the structure in the low-pressure range. Depending on the chosen 
chemical composition, the Fe moments either survive or vanish across the structural 
collapse. In the case of BaCsFe$_4$As$_4$, the local Fe moments are suppressed even before 
the formation of As-As bonds.

Based on our calculations, the critical value for the As-As distance leading to a structural 
collapse varies between (2.8\,--\,3.0)\,\AA{}, depending on the spacer cation size. The initial 
As-As distance at zero pressure determines then the critical pressure necessary for a structural 
transition. Finding ways of bringing the 1144 systems closer to the critical As-As separation 
already at zero pressure will eventually allow the observation of half-collapsed phases at 
ambient pressure in this class of materials. For instance, the critical pressures at which the 
half collapses happen can be significantly reduced in uniaxially compressed systems as has been 
already observed in 122 systems,\cite{Prokes2010,Tomic2012,Tomic2013} making the observation of 
this structural transition, in principle, possible for some of the explored compounds. Also, 
study of collapsed tetragonal phases in other 1144 systems based on tri- and monovalent spacer 
cations that were newly suggested to be stable at ambient pressure,\cite{Song2017,Song2018} 
e.g. LaKFe$_4$As$_4$ and EuKFe$_4$As$_4$, might be another promising research direction.

For the EuRbFe$_4$As$_4$ and EuCsFe$_4$As$_4$ systems, the character of the half-collapsed 
transition and the critical pressures are similar to the Sr-based systems which is explained 
by almost identical ionic radii of Eu and Sr 2+ cations. The Eu magnetism is stable across both 
collapse transitions, with the \textit{4f} moments staying close to the expected value of 
$7\,\mu_\mathrm{B}$. Very recent experimental observations\cite{Jackson2018} seem to confirm 
the predicted pressure trends.

In view of the novel magnetism discovered recently in electron-doped
\CaK{},\cite{Meier2018} the rest of the 1144 family is worth studying in
terms of their magnetic proper\-ties. CaRbFe$_4$As$_4$ and CaCsFe$_4$As$_4$
 have the largest As height asymmetry and are therefore promising
candidates for a spin-vortex magnetic crystal with a high ordering temperature.
Search for other 1144 materials with a larger difference in the two cation
sizes, leading to a stronger asymmetry between the As sites, appears to be a
natural next step. In contrast, 1144 systems with bigger cations, such as 
BaCsFe$_4$As$_4$, have a more symmetric structure and are likely to be closer 
to stripe order than more asymmetric 1144 compounds. The motivation to look for 
further spin-vortex ordered Fe-based systems is the recently proposed relation of 
this magnetic order to the strong unconventional superconductivity in 1144 systems. 
Furthermore, this is a first example of a non-collinear double-$Q$ magnetic order 
in iron pnictides and presents an important part of their general phase diagram.\cite{Meier2018}

\textit{Acknowledgments.--} This work was financially supported by DFG
Sonderforschungsbereich TRR 49. Work at Ames Lab (PCC) was supported by the U.S. 
Department of Energy, Office of Basic Energy Science, Division of Materials Sciences 
and Engineering. Ames Laboratory is operated for the U.S. Department of Energy by 
Iowa State University under Contract No. DE-AC02-07CH11358. Research (VB) was funded 
in part by the Gordon and Betty Moore Foundation’s EPiQS Initiative through Grant 
GBMF4411. This work was initiated as part of the Humboldt Prize for PCC. 
The computer time was allotted by the centre for supercomputing (CSC) in Frankfurt 
and by the computer center of Goethe University. VB and RV acknowledge discussions 
with Peter Hirschfeld and James Hamlin. Parts of some figures have been produced 
with \textsc{VESTA3} (Ref.~\onlinecite{Vesta}).

\bibliographystyle{prb-titles} \bibliography{./paper}

\begin{thebibliography}{10}
\providecommand{\bibAnnoteFile}[1]{%
  \IfFileExists{#1}{\begin{quotation}\noindent\textsc{Key:} #1\\
  \textsc{Annotation:}\ \input{#1}\end{quotation}}{}}
\providecommand{\bibAnnote}[2]{%
  \begin{quotation}\noindent\textsc{Key:} #1\\
  \textsc{Annotation:}\ #2\end{quotation}}
\providecommand{\bibinfo}[2]{#2}

\bibitem{Kreyssig2008}
\bibinfo{author}{A.~Kreyssig}, \bibinfo{author}{M.~A. Green},
  \bibinfo{author}{Y.~Lee}, \bibinfo{author}{G.~D. Samolyuk},
  \bibinfo{author}{P.~Zajdel}, \bibinfo{author}{J.~W. Lynn},
  \bibinfo{author}{S.~L. Bud'ko}, \bibinfo{author}{M.~S. Torikachvili},
  \bibinfo{author}{N.~Ni}, \bibinfo{author}{S.~Nandi}, \bibinfo{author}{J.~B.
  Le\~ao}, \bibinfo{author}{S.~J. Poulton}, \bibinfo{author}{D.~N. Argyriou},
  \bibinfo{author}{B.~N. Harmon}, \bibinfo{author}{R.~J. McQueeney},
  \bibinfo{author}{P.~C. Canfield}, and \bibinfo{author}{A.~I. Goldman},
  \bibinfo{title}{Pressure-induced volume-collapsed tetragonal phase of
  ${\text{CaFe}}_{2}{\text{As}}_{2}$ as seen via neutron scattering},
  \bibinfo{journal}{Phys. Rev. B} \textbf{\bibinfo{volume}{78}},
  \bibinfo{pages}{184517} (\bibinfo{year}{2008}).
\bibAnnoteFile{Kreyssig2008}

\bibitem{Canfield2010}
\bibinfo{author}{P.~C. Canfield} and \bibinfo{author}{S.~L. Bud'ko},
  \bibinfo{title}{FeAs-Based Superconductivity: A Case Study of the Effects of
  Transition Metal Doping on BaFe$_2$As$_2$}, \bibinfo{journal}{Annual Review
  of Condensed Matter Physics} \textbf{\bibinfo{volume}{1}},
  \bibinfo{pages}{27} (\bibinfo{year}{2010}).
\bibAnnoteFile{Canfield2010}

\bibitem{Yildirim2009}
\bibinfo{author}{T.~Yildirim}, \bibinfo{title}{Strong Coupling of the Fe-Spin
  State and the As-As Hybridization in Iron-Pnictide Superconductors from
  First-Principle Calculations}, \bibinfo{journal}{Phys. Rev. Lett.}
  \textbf{\bibinfo{volume}{102}}, \bibinfo{pages}{037003}
  (\bibinfo{year}{2009}).
\bibAnnoteFile{Yildirim2009}

\bibitem{Kaluarachchi2017}
\bibinfo{author}{U.~S. Kaluarachchi}, \bibinfo{author}{V.~Taufour},
  \bibinfo{author}{A.~Sapkota}, \bibinfo{author}{V.~Borisov},
  \bibinfo{author}{T.~Kong}, \bibinfo{author}{W.~R. Meier},
  \bibinfo{author}{K.~Kothapalli}, \bibinfo{author}{B.~G. Ueland},
  \bibinfo{author}{A.~Kreyssig}, \bibinfo{author}{R.~Valent\'i},
  \bibinfo{author}{R.~J. McQueeney}, \bibinfo{author}{A.~I. Goldman},
  \bibinfo{author}{S.~L. Bud'ko}, and \bibinfo{author}{P.~C. Canfield},
  \bibinfo{title}{Pressure-induced half-collapsed-tetragonal phase in
  CaKFe$_4$As$_4$}, \bibinfo{journal}{Phys. Rev. B}
  \textbf{\bibinfo{volume}{96}}, \bibinfo{pages}{140501(R)}
  (\bibinfo{year}{2017}).
\bibAnnoteFile{Kaluarachchi2017}

\bibitem{Iyo2016}
\bibinfo{author}{A.~Iyo}, \bibinfo{author}{K.~Kawashima},
  \bibinfo{author}{T.~Kinjo}, \bibinfo{author}{T.~Nishio},
  \bibinfo{author}{S.~Ishida}, \bibinfo{author}{H.~Fujihisa},
  \bibinfo{author}{Y.~Gotoh}, \bibinfo{author}{K.~Kihou},
  \bibinfo{author}{H.~Eisaki}, and \bibinfo{author}{Y.~Yoshida},
  \bibinfo{title}{New-Structure-Type Fe-Based Superconductors:
  Ca\textit{A}Fe$_4$As$_4$ (\textit{A} = K, Rb, Cs) and
  Sr\textit{A}Fe$_4$As$_4$ (\textit{A} = Rb, Cs)}, \bibinfo{journal}{J. Am.
  Chem. Soc.} \textbf{\bibinfo{volume}{138 (10)}}, \bibinfo{pages}{3410}
  (\bibinfo{year}{2016}).
\bibAnnoteFile{Iyo2016}

\bibitem{Meier2016}
\bibinfo{author}{W.~R. Meier}, \bibinfo{author}{T.~Kong},
  \bibinfo{author}{U.~S. Kaluarachchi}, \bibinfo{author}{V.~Taufour},
  \bibinfo{author}{N.~H. Jo}, \bibinfo{author}{G.~Drachuck},
  \bibinfo{author}{A.~E. B\"ohmer}, \bibinfo{author}{S.~M. Saunders},
  \bibinfo{author}{A.~Sapkota}, \bibinfo{author}{A.~Kreyssig},
  \bibinfo{author}{M.~A. Tanatar}, \bibinfo{author}{R.~Prozorov},
  \bibinfo{author}{A.~I. Goldman}, \bibinfo{author}{F.~F. Balakirev},
  \bibinfo{author}{A.~Gurevich}, \bibinfo{author}{S.~L. Bud'ko}, and
  \bibinfo{author}{P.~C. Canfield}, \bibinfo{title}{Anisotropic thermodynamic
  and transport properties of single-crystalline CaKFe$_4$As$_4$},
  \bibinfo{journal}{Phys. Rev. B} \textbf{\bibinfo{volume}{94}},
  \bibinfo{pages}{064501} (\bibinfo{year}{2016}).
\bibAnnoteFile{Meier2016}

\bibitem{Budko2017}
\bibinfo{author}{S.~L. Bud'ko}, \bibinfo{author}{T.~Kong},
  \bibinfo{author}{W.~R. Meier}, \bibinfo{author}{X.~Ma}, and
  \bibinfo{author}{P.~C. Canfield}, \bibinfo{title}{$^{57}$Fe M\"ossbauer study
  of stoichiometric iron-based superconductor CaKFe$_4$As$_4$: a comparison to
  KFe$_2$As$_2$ and CaFe$_2$As$_2$}, \bibinfo{journal}{Philosophical Magazine}
  \textbf{\bibinfo{volume}{97}}, \bibinfo{pages}{2689} (\bibinfo{year}{2017}).
\bibAnnoteFile{Budko2017}

\bibitem{Meier2018}
\bibinfo{author}{W.~R. Meier}, \bibinfo{author}{Q.-P. Ding},
  \bibinfo{author}{A.~Kreyssig}, \bibinfo{author}{S.~L. Bud'ko},
  \bibinfo{author}{A.~Sapkota}, \bibinfo{author}{K.~Kothapalli},
  \bibinfo{author}{V.~Borisov}, \bibinfo{author}{R.~Valent\'i},
  \bibinfo{author}{C.~D. Batista}, \bibinfo{author}{P.~P. Orth},
  \bibinfo{author}{R.~M. Fernandes}, \bibinfo{author}{A.~I. Goldman},
  \bibinfo{author}{Y.~Furukawa}, \bibinfo{author}{A.~E. B\"ohmer}, and
  \bibinfo{author}{P.~C. Canfield}, \bibinfo{title}{Hedgehog spin-vortex
  crystal stabilized in a hole-doped iron-based superconductor},
  \bibinfo{journal}{npj Quantum Materials} \textbf{\bibinfo{volume}{3}},
  \bibinfo{pages}{5} (\bibinfo{year}{2018}).
\bibAnnoteFile{Meier2018}

\bibitem{Shannon1976}
\bibinfo{author}{R.~D. Shannon}, \bibinfo{title}{Revised Effective Ionic Radii
  and Systematic Studies of Interatomic Distances in Halides and
  Chalcogenides}, \bibinfo{journal}{Acta Cryst.} \textbf{\bibinfo{volume}{A
  32}}, \bibinfo{pages}{751} (\bibinfo{year}{1976}).
\bibAnnoteFile{Shannon1976}

\bibitem{Kawashima2016}
\bibinfo{author}{K.~Kawashima}, \bibinfo{author}{T.~Kinjo},
  \bibinfo{author}{T.~Nishio}, \bibinfo{author}{S.~Ishida},
  \bibinfo{author}{H.~Fujihisa}, \bibinfo{author}{Y.~Gotoh},
  \bibinfo{author}{K.~Kihou}, \bibinfo{author}{H.~Eisaki},
  \bibinfo{author}{Y.~Yoshida}, and \bibinfo{author}{A.~Iyo},
  \bibinfo{title}{Superconductivity in Fe-Based Compound Eu$A$Fe$_4$As$_4$ (A =
  Rb and Cs)}, \bibinfo{journal}{J. Phys. Soc. Jpn.}
  \textbf{\bibinfo{volume}{85}}, \bibinfo{pages}{064710}
  (\bibinfo{year}{2016}).
\bibAnnoteFile{Kawashima2016}

\bibitem{Liu2016}
\bibinfo{author}{Y.~Liu}, \bibinfo{author}{Y.-B. Liu},
  \bibinfo{author}{Q.~Chen}, \bibinfo{author}{Z.-T. Tang},
  \bibinfo{author}{W.-H. Jiao}, \bibinfo{author}{Q.~Tao},
  \bibinfo{author}{Z.-A. Xu}, and \bibinfo{author}{G.-H. Cao},
  \bibinfo{title}{A new ferromagnetic superconductor: CsEuFe$_4$As$_4$},
  \bibinfo{journal}{Sci. Bull.} \textbf{\bibinfo{volume}{61}},
  \bibinfo{pages}{1213} (\bibinfo{year}{2016}).
\bibAnnoteFile{Liu2016}

\bibitem{Liu2017}
\bibinfo{author}{Y.~Liu}, \bibinfo{author}{Y.-B. Liu}, \bibinfo{author}{Y.-L.
  Yu}, \bibinfo{author}{Q.~Tao}, \bibinfo{author}{C.-M. Feng}, and
  \bibinfo{author}{G.-H. Cao},
  \bibinfo{title}{RbEu(Fe$_{1-x}$Ni$_x$)$_4$As$_4$: From a ferromagnetic
  superconductor to a superconducting ferromagnet}, \bibinfo{journal}{Phys.
  Rev. B} \textbf{\bibinfo{volume}{96}}, \bibinfo{pages}{224510}
  (\bibinfo{year}{2017}).
\bibAnnoteFile{Liu2017}

\bibitem{Jackson2018}
\bibinfo{author}{D.~E. Jackson}, \bibinfo{author}{D.~VanGennep},
  \bibinfo{author}{W.~Bi}, \bibinfo{author}{D.~Zhang},
  \bibinfo{author}{P.~Materne}, \bibinfo{author}{Y.~Liu},
  \bibinfo{author}{G.-H. Cao}, \bibinfo{author}{S.~T. Weir},
  \bibinfo{author}{Y.~K. Vohra}, and \bibinfo{author}{J.~J. Hamlin},
  \bibinfo{title}{Superconducting and magnetic phase diagram of
  RbEuFe$_4$As$_4$ and CsEuFe$_4$As$_4$ at high pressure},
  \bibinfo{journal}{arXiv:1805.09288}  (\bibinfo{year}{2018}).
\bibAnnoteFile{Jackson2018}

\bibitem{Zapf2017}
\bibinfo{author}{S.~Zapf} and \bibinfo{author}{M.~Dressel},
  \bibinfo{title}{Europium-based iron pnictides: a unique laboratory for
  magnetism, superconductivity and structural effects},
  \bibinfo{journal}{Reports on Progress in Physics}
  \textbf{\bibinfo{volume}{80}}, \bibinfo{pages}{016501}
  (\bibinfo{year}{2017}).
\bibAnnoteFile{Zapf2017}

\bibitem{Sypek2017}
\bibinfo{author}{J.~T. Sypek}, \bibinfo{author}{H.~Yu}, \bibinfo{author}{K.~J.
  Dusoe}, \bibinfo{author}{G.~Drachuck}, \bibinfo{author}{H.~Patel},
  \bibinfo{author}{A.~M. Giroux}, \bibinfo{author}{A.~I. Goldman},
  \bibinfo{author}{A.~Kreyssig}, \bibinfo{author}{P.~C. Canfield},
  \bibinfo{author}{S.~L. Bud'ko}, \bibinfo{author}{C.~R. Weinberger}, and
  \bibinfo{author}{S.-W. Lee}, \bibinfo{title}{Superelasticity and cryogenic
  linear shape memory effects of CaFe$_2$As$_2$}, \bibinfo{journal}{Nat.
  Commun.} \textbf{\bibinfo{volume}{8}} (\bibinfo{year}{2017}).
\bibAnnoteFile{Sypek2017}

\bibitem{Note1}
\bibinfo{note}{We note that the exact crystal structure of BaCsFe$_4$As$_4$ is
  not completely clarified and this material can probably exist both in the 122
  and in the 1144 forms due to a smaller difference in the cation radii,
  compared to other 1144 systems. Nevertheless, we include the 1144 phase of
  BaCsFe$_4$As$_4$ in our simulations, in order to reveal general trends in the
  1144 family of iron pnictides.}
\bibAnnoteFile{Note1}

\bibitem{Bloechl1994}
\bibinfo{author}{P.~E. Bl\"ochl}, \bibinfo{title}{Projector augmented-wave
  method}, \bibinfo{journal}{Phys. Rev. B} \textbf{\bibinfo{volume}{50}},
  \bibinfo{pages}{17953} (\bibinfo{year}{1994}).
\bibAnnoteFile{Bloechl1994}

\bibitem{Kresse1999}
\bibinfo{author}{G.~Kresse} and \bibinfo{author}{D.~Joubert},
  \bibinfo{title}{From ultrasoft pseudopotentials to the projector
  augmented-wave method}, \bibinfo{journal}{Phys. Rev. B}
  \textbf{\bibinfo{volume}{59}}, \bibinfo{pages}{1758} (\bibinfo{year}{1999}).
\bibAnnoteFile{Kresse1999}

\bibitem{Kresse1993}
\bibinfo{author}{G.~Kresse} and \bibinfo{author}{J.~Hafner},
  \bibinfo{title}{\textit{Ab initio} molecular dynamics for liquid metals},
  \bibinfo{journal}{Phys. Rev. B (Rapid Communications)}
  \textbf{\bibinfo{volume}{47}}, \bibinfo{pages}{558} (\bibinfo{year}{1993}).
\bibAnnoteFile{Kresse1993}

\bibitem{Kresse1996}
\bibinfo{author}{G.~Kresse} and \bibinfo{author}{J.~Furthm\"uller},
  \bibinfo{title}{Efficient iterative schemes for \textit{ab initio}
  total-energy calculations using a plane-wave basis set},
  \bibinfo{journal}{Phys. Rev. B} \textbf{\bibinfo{volume}{54}},
  \bibinfo{pages}{11169} (\bibinfo{year}{1996}).
\bibAnnoteFile{Kresse1996}

\bibitem{Kresse1996b}
\bibinfo{author}{G.~Kresse} and \bibinfo{author}{J.~Furthm\"uller},
  \bibinfo{title}{Efficiency of ab-initio total energy calculations for metals
  and semiconductors using a plane-wave basis set}, \bibinfo{journal}{Comput.
  Mater. Sci.} \textbf{\bibinfo{volume}{6}}, \bibinfo{pages}{15}
  (\bibinfo{year}{1996}).
\bibAnnoteFile{Kresse1996b}

\bibitem{Perdew1996}
\bibinfo{author}{J.~P. Perdew}, \bibinfo{author}{K.~Burke}, and
  \bibinfo{author}{M.~Ernzerhof}, \bibinfo{title}{Generalized gradient
  approximation made simple}, \bibinfo{journal}{Phys. Rev. Lett.}
  \textbf{\bibinfo{volume}{77}}, \bibinfo{pages}{3865} (\bibinfo{year}{1996}).
\bibAnnoteFile{Perdew1996}

\bibitem{Dudarev1998}
\bibinfo{author}{S.~L. Dudarev}, \bibinfo{author}{G.~A. Botton},
  \bibinfo{author}{S.~Y. Savrasov}, \bibinfo{author}{C.~J. Humphreys}, and
  \bibinfo{author}{A.~P. Sutton}, \bibinfo{title}{Electron-energy-loss spectra
  and the structural stability of nickel oxide: An {LSDA+U} study},
  \bibinfo{journal}{Phys. Rev. B} \textbf{\bibinfo{volume}{57}},
  \bibinfo{pages}{1505} (\bibinfo{year}{1998}).
\bibAnnoteFile{Dudarev1998}

\bibitem{Liechtenstein1995}
\bibinfo{author}{A.~I. Liechtenstein}, \bibinfo{author}{V.~I. Anisimov}, and
  \bibinfo{author}{J.~Zaanen}, \bibinfo{title}{Density-functional theory and
  strong interactions: Orbital ordering in Mott-Hubbard insulators},
  \bibinfo{journal}{Phys. Rev. B} \textbf{\bibinfo{volume}{52}},
  \bibinfo{pages}{R5467} (\bibinfo{year}{1995}).
\bibAnnoteFile{Liechtenstein1995}

\bibitem{Cui2017}
\bibinfo{author}{J.~Cui}, \bibinfo{author}{Q.-P. Ding}, \bibinfo{author}{W.~R.
  Meier}, \bibinfo{author}{A.~E. B\"ohmer}, \bibinfo{author}{T.~Kong},
  \bibinfo{author}{V.~Borisov}, \bibinfo{author}{Y.~Lee},
  \bibinfo{author}{S.~L. Bud'ko}, \bibinfo{author}{R.~Valent\'i},
  \bibinfo{author}{P.~C. Canfield}, and \bibinfo{author}{Y.~Furukawa},
  \bibinfo{title}{Magnetic fluctuations and superconducting properties of
  CaKFe$_4$As$_4$ studied by $^{75}$As NMR}, \bibinfo{journal}{Phys. Rev. B}
  \textbf{\bibinfo{volume}{96}}, \bibinfo{pages}{104512}
  (\bibinfo{year}{2017}).
\bibAnnoteFile{Cui2017}

\bibitem{Zhang2009}
\bibinfo{author}{Y.-Z. Zhang}, \bibinfo{author}{H.~C. Kandpal},
  \bibinfo{author}{I.~Opahle}, \bibinfo{author}{H.~O. Jeschke}, and
  \bibinfo{author}{R.~Valent\'{\i}}, \bibinfo{title}{Microscopic origin of
  pressure-induced phase transitions in the iron pnictide superconductors
  $A{\text{Fe}}_{2}{\text{As}}_{2}$: An ab initio molecular dynamics study},
  \bibinfo{journal}{Phys. Rev. B} \textbf{\bibinfo{volume}{80}},
  \bibinfo{pages}{094530} (\bibinfo{year}{2009}).
\bibAnnoteFile{Zhang2009}

\bibitem{Colonna2011}
\bibinfo{author}{N.~Colonna}, \bibinfo{author}{G.~Profeta},
  \bibinfo{author}{A.~Continenza}, and \bibinfo{author}{S.~Massidda},
  \bibinfo{title}{Structural and magnetic properties of CaFe$_2$As$_2$ and
  BaFe$_2$As$_2$ from first-principles density functional theory},
  \bibinfo{journal}{Phys. Rev. B} \textbf{\bibinfo{volume}{83}},
  \bibinfo{pages}{094529} (\bibinfo{year}{2011}).
\bibAnnoteFile{Colonna2011}

\bibitem{Tomic2012}
\bibinfo{author}{M.~Tomi\'c}, \bibinfo{author}{R.~Valent\'i}, and
  \bibinfo{author}{H.~O. Jeschke}, \bibinfo{title}{Uniaxial versus hydrostatic
  pressure-induced phase transitions in CaFe$_2$As$_2$ and BaFe$_2$As$_2$},
  \bibinfo{journal}{Phys. Rev. B} \textbf{\bibinfo{volume}{85}},
  \bibinfo{pages}{094105} (\bibinfo{year}{2012}).
\bibAnnoteFile{Tomic2012}

\bibitem{Dhaka2014}
\bibinfo{author}{R.~S. Dhaka}, \bibinfo{author}{R.~Jiang},
  \bibinfo{author}{S.~Ran}, \bibinfo{author}{S.~L. Bud'ko},
  \bibinfo{author}{P.~C. Canfield}, \bibinfo{author}{B.~N. Harmon},
  \bibinfo{author}{A.~Kaminski}, \bibinfo{author}{M.~Tomi\'c},
  \bibinfo{author}{R.~Valent\'i}, and \bibinfo{author}{Y.~Lee},
  \bibinfo{title}{Dramatic changes in the electronic structure upon transition
  to the collapsed tetragonal phase in CaFe$_2$As$_2$}, \bibinfo{journal}{Phys.
  Rev. B} \textbf{\bibinfo{volume}{89}}, \bibinfo{pages}{020511}
  (\bibinfo{year}{2014}).
\bibAnnoteFile{Dhaka2014}

\bibitem{Budko2016}
\bibinfo{author}{S.~L. Bud'ko}, \bibinfo{author}{X.~Ma},
  \bibinfo{author}{M.~Tomi\ifmmode~\acute{c}\else \'{c}\fi{}},
  \bibinfo{author}{S.~Ran}, \bibinfo{author}{R.~Valent\'{\i}}, and
  \bibinfo{author}{P.~C. Canfield}, \bibinfo{title}{Transition to collapsed
  tetragonal phase in ${\mathrm{CaFe}}_{2}{\mathrm{As}}_{2}$ single crystals as
  seen by $^{57}\mathrm{Fe}$ M\"ossbauer spectroscopy}, \bibinfo{journal}{Phys.
  Rev. B} \textbf{\bibinfo{volume}{93}}, \bibinfo{pages}{024516}
  (\bibinfo{year}{2016}).
\bibAnnoteFile{Budko2016}

\bibitem{FPLO}
\bibinfo{author}{K.~Koepernik} and \bibinfo{author}{H.~Eschrig},
  \bibinfo{title}{Full-potential nonorthogonal local-orbital minimum-basis
  band-structure scheme}, \bibinfo{journal}{Phys. Rev. B}
  \textbf{\bibinfo{volume}{59}}, \bibinfo{pages}{1743} (\bibinfo{year}{1999}).
\bibAnnoteFile{FPLO}

\bibitem{Kasinathan2011}
\bibinfo{author}{D.~Kasinathan}, \bibinfo{author}{M.~Schmitt},
  \bibinfo{author}{K.~Koepernik}, \bibinfo{author}{A.~Ormeci},
  \bibinfo{author}{K.~Meier}, \bibinfo{author}{U.~Schwarz},
  \bibinfo{author}{M.~Hanfland}, \bibinfo{author}{C.~Geibel},
  \bibinfo{author}{Y.~Grin}, \bibinfo{author}{A.~Leithe-Jasper}, and
  \bibinfo{author}{H.~Rosner}, \bibinfo{title}{Symmetry-preserving lattice
  collapse in tetragonal SrFe${}_{2\ensuremath{-}x}$Ru${}_{x}$As${}_{2}$
  ($x=0,0.2$): A combined experimental and theoretical study},
  \bibinfo{journal}{Phys. Rev. B} \textbf{\bibinfo{volume}{84}},
  \bibinfo{pages}{054509} (\bibinfo{year}{2011}).
\bibAnnoteFile{Kasinathan2011}

\bibitem{Diehl2014}
\bibinfo{author}{J.~Diehl}, \bibinfo{author}{S.~Backes},
  \bibinfo{author}{D.~Guterding}, \bibinfo{author}{H.~O. Jeschke}, and
  \bibinfo{author}{R.~Valent\'i}, \bibinfo{title}{Correlation effects in the
  tetragonal and collapsed-tetragonal phase of CaFe$_2$As$_2$},
  \bibinfo{journal}{Phys. Rev. B} \textbf{\bibinfo{volume}{90}},
  \bibinfo{pages}{085110} (\bibinfo{year}{2014}).
\bibAnnoteFile{Diehl2014}

\bibitem{Note2}
\bibinfo{note}{Please note that in our simulations we didn't consider more
  complex cases of Eu magnetism than ferromagnetism.}
\bibAnnoteFile{Note2}

\bibitem{Mittal2011}
\bibinfo{author}{R.~Mittal}, \bibinfo{author}{S.~K. Mishra},
  \bibinfo{author}{S.~L. Chaplot}, \bibinfo{author}{S.~V. Ovsyannikov},
  \bibinfo{author}{E.~Greenberg}, \bibinfo{author}{D.~M. Trots},
  \bibinfo{author}{L.~Dubrovinsky}, \bibinfo{author}{Y.~Su},
  \bibinfo{author}{T.~Brueckel}, \bibinfo{author}{S.~Matsuishi},
  \bibinfo{author}{H.~Hosono}, and \bibinfo{author}{G.~Garbarino},
  \bibinfo{title}{Ambient- and low-temperature synchrotron x-ray diffraction
  study of BaFe$_2$As$_2$ and CaFe$_2$As$_2$ at high pressures up to 56 GPa},
  \bibinfo{journal}{Phys. Rev. B} \textbf{\bibinfo{volume}{83}},
  \bibinfo{pages}{054503} (\bibinfo{year}{2011}).
\bibAnnoteFile{Mittal2011}

\bibitem{Tomic2013}
\bibinfo{author}{M.~Tomi\'{c}}, \bibinfo{author}{H.~O. Jeschke},
  \bibinfo{author}{R.~M. Fernandes}, and \bibinfo{author}{R.~Valent\'i},
  \bibinfo{title}{In-plane uniaxial stress effects on the structural and
  electronic properties of BaFe${}_{2}$As${}_{2}$ and CaFe${}_{2}$As${}_{2}$},
  \bibinfo{journal}{Phys. Rev. B} \textbf{\bibinfo{volume}{87}},
  \bibinfo{pages}{174503} (\bibinfo{year}{2013}).
\bibAnnoteFile{Tomic2013}

\bibitem{Guterding2017}
\bibinfo{author}{D.~Guterding}, \bibinfo{author}{S.~Backes},
  \bibinfo{author}{M.~Tomi\'c}, \bibinfo{author}{H.~O. Jeschke}, and
  \bibinfo{author}{R.~Valent\'i}, \bibinfo{title}{Ab initio perspective on
  structural and electronic properties of iron‐based superconductors},
  \bibinfo{journal}{physica status solidi (b)} \textbf{\bibinfo{volume}{254}},
  \bibinfo{pages}{1600164}.
\bibAnnoteFile{Guterding2017}

\bibitem{Uhoya2010}
\bibinfo{author}{W.~Uhoya}, \bibinfo{author}{G.~Tsoi}, \bibinfo{author}{Y.~K.
  Vohra}, \bibinfo{author}{M.~A. McGuire}, \bibinfo{author}{A.~S. Sefat},
  \bibinfo{author}{B.~C. Sales}, \bibinfo{author}{D.~Mandrus}, and
  \bibinfo{author}{S.~T. Weir}, \bibinfo{title}{Anomalous compressibility
  effects and superconductivity of EuFe$_2$As$_2$ under high pressures},
  \bibinfo{journal}{Journal of Physics: Condensed Matter}
  \textbf{\bibinfo{volume}{22}}, \bibinfo{pages}{292202}
  (\bibinfo{year}{2010}).
\bibAnnoteFile{Uhoya2010}

\bibitem{Prokes2010}
\bibinfo{author}{K.~Proke\ifmmode~\check{s}\else \v{s}\fi{}},
  \bibinfo{author}{A.~Kreyssig}, \bibinfo{author}{B.~Ouladdiaf},
  \bibinfo{author}{D.~K. Pratt}, \bibinfo{author}{N.~Ni},
  \bibinfo{author}{S.~L. Bud'ko}, \bibinfo{author}{P.~C. Canfield},
  \bibinfo{author}{R.~J. McQueeney}, \bibinfo{author}{D.~N. Argyriou}, and
  \bibinfo{author}{A.~I. Goldman}, \bibinfo{title}{Evidence from neutron
  diffraction for superconductivity in the stabilized tetragonal phase of
  ${\text{CaFe}}_{2}{\text{As}}_{2}$ under uniaxial pressure},
  \bibinfo{journal}{Phys. Rev. B} \textbf{\bibinfo{volume}{81}},
  \bibinfo{pages}{180506} (\bibinfo{year}{2010}).
\bibAnnoteFile{Prokes2010}

\bibitem{Song2017}
\bibinfo{author}{B.~Q. Song}, \bibinfo{author}{M.~C. Nguyen},
  \bibinfo{author}{C.~Z. Wang}, \bibinfo{author}{P.~C. Canfield}, and
  \bibinfo{author}{K.~M. Ho}, \bibinfo{title}{Whether it is possible to
  stabilize the 1144-phase pnictides with tri-valence cations?},
  \bibinfo{journal}{arXiv:1710.01868}  (\bibinfo{year}{2017}).
\bibAnnoteFile{Song2017}

\bibitem{Song2018}
\bibinfo{author}{B.~Q. Song}, \bibinfo{author}{M.~C. Nguyen},
  \bibinfo{author}{C.~Z. Wang}, and \bibinfo{author}{K.~M. Ho},
  \bibinfo{title}{Stability of the 1144 phase in iron pnictides},
  \bibinfo{journal}{Phys. Rev. B} \textbf{\bibinfo{volume}{97}},
  \bibinfo{pages}{094105} (\bibinfo{year}{2018}).
\bibAnnoteFile{Song2018}

\bibitem{Vesta}
\bibinfo{author}{K.~Momma} and \bibinfo{author}{F.~Izumi},
  \bibinfo{title}{{VESTA3 for three-dimensional visualization of crystal,
  volumetric and morphology data}}, \bibinfo{journal}{J. Appl. Cryst.}
  \textbf{\bibinfo{volume}{44}}, \bibinfo{pages}{1272} (\bibinfo{year}{2011}).
\bibAnnoteFile{Vesta}

\end{thebibliography}

\newpage

\appendix

\section{Role of local Fe moments}

\begin{figure}[H]
  \includegraphics[width = 0.99\columnwidth]{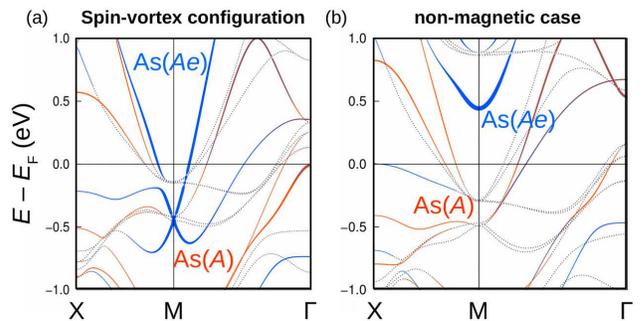}
  \caption{Comparison of the As-$4p_z$ orbitals for CaRbFe$_4$As$_4$ structures relaxed at zero pressure (a) using the spin-vortex Fe order or (b) non-magnetically. In the latter case, the As-As bonds across Ca are already formed, as opposed to the former structure, which emphasizes the importance of spin-vortex fluctuations in 1144 systems. For both relaxed geometries, the band structure is non-spin-polarized. The color code is the same as in Fig.~\ref{f:collapse_criterium}.} \label{f:zero_pressure_As_bands}
\end{figure}

As emphasized in the main text, the presence of local Fe moments is a necessary 
ingredient of our simulations, since otherwise the predicted structures would 
have a largely underestimated $c$ lattice parameter, which, in case of CaRbFe$_4$As$_4$, 
even leads to the formation of As-As bonds near Ca already at zero pressure, confirmed 
by the band structure analysis (Fig.~\ref{f:zero_pressure_As_bands}).
We have also performed some test calculations with Fe magnetic orders other 
than spin-vortex and found that the best agreement with structural parameters 
from experiment is achieved assuming a spin-vortex arrangement of Fe moments.

\newpage

\section{Accuracy of structure optimization}

\begin{figure}[H]
  \includegraphics[width = 0.99\columnwidth]{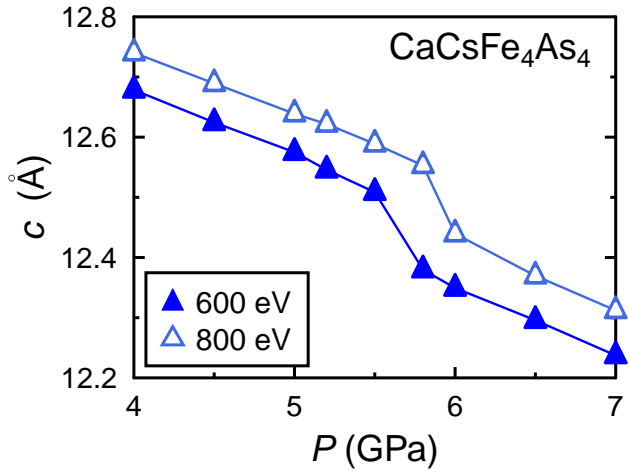}\vspace{-15pt}
  \caption{Comparison of the calculated pressure-dependent $c$-lattice parameter of CaCsFe$_4$As$_4$ for two different energy cutoff values 600~eV and 800~eV.} \label{f:energy_cutoff_effect}
\end{figure}

Regarding the accuracy of our pressure simulations, as mentioned in the main text 
and illustrated in Fig.~\ref{f:energy_cutoff_effect}, the increase of the 
energy cutoff from 600~eV to 800~eV leads to rather small changes in the lattice 
parameters, which do not affect the general trends in the 1144 family.

\section{Structure-evolution under pressure}

\begin{figure*}
  \includegraphics[width = 0.99\textwidth]{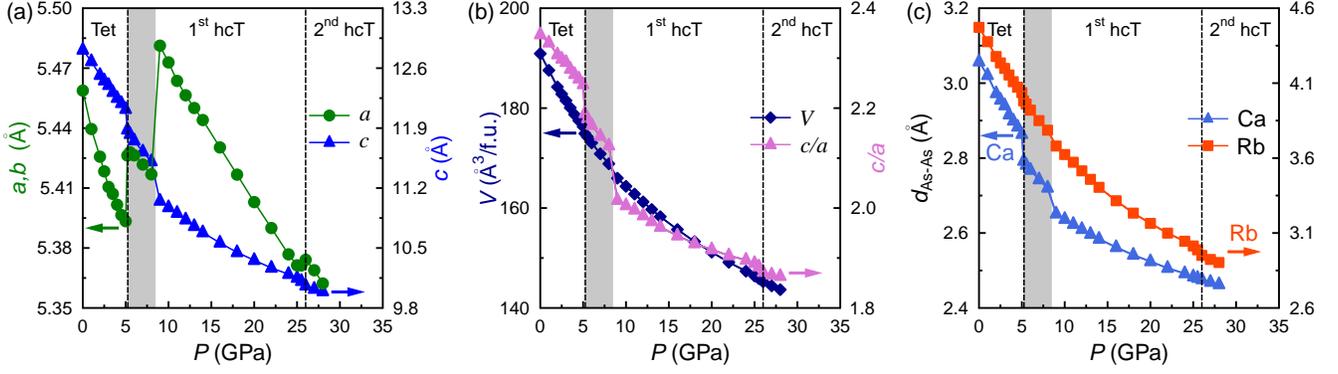}\vspace{-5pt}
  \caption{Pressure evolution of (a) lattice parameters $a=b$ and $c$, (b) volume and $c/a$ ratio and (c) As-As distances across both hcT transitions for CaRbFe$_4$As$_4$. The critical pressures of the two half-collapsed transitions are marked by vertical dashed lines and the pressure range between the first hcT and the subsequent collapse of Fe moments is indicated by shading.} \label{f:CaRb_structure_vs_pressure}
\end{figure*}

\begin{figure*}
  \includegraphics[width = 0.99\textwidth]{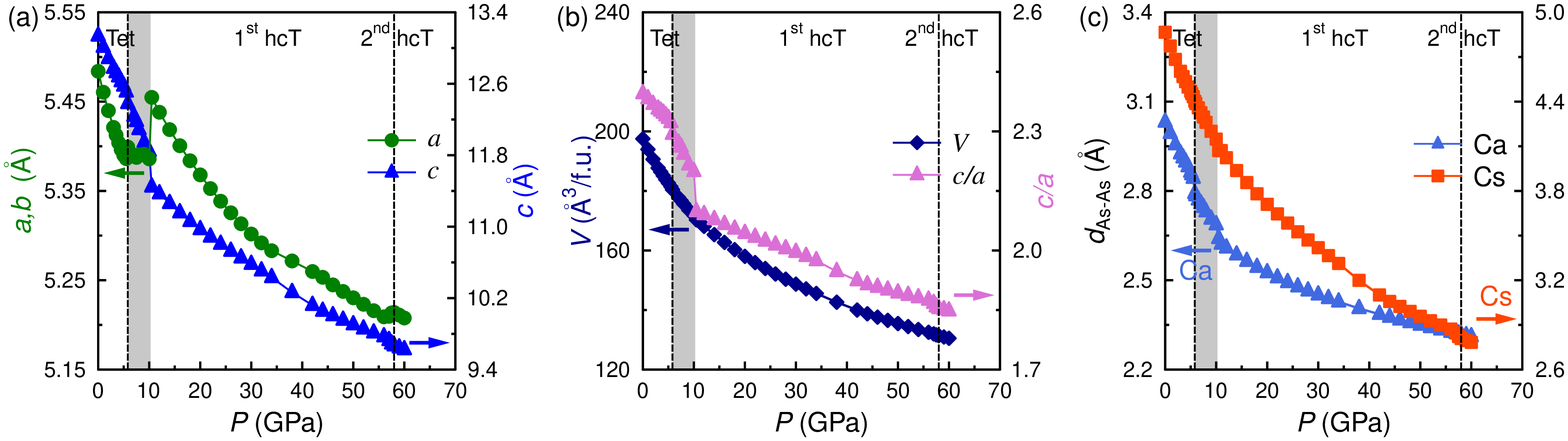}\vspace{-5pt}
  \caption{Pressure evolution of (a) lattice parameters $a=b$ and $c$, (b) volume and $c/a$ ratio and (c) As-As distances across both hcT transitions for CaCsFe$_4$As$_4$. The critical pressures of the two half-collapsed transitions are marked by vertical dashed lines and the pressure range between the first hcT and the subsequent collapse of Fe moments is indicated by shading.} \label{f:CaCs_structure_vs_pressure}
\end{figure*}

\begin{figure*}
  \includegraphics[width = 0.99\textwidth]{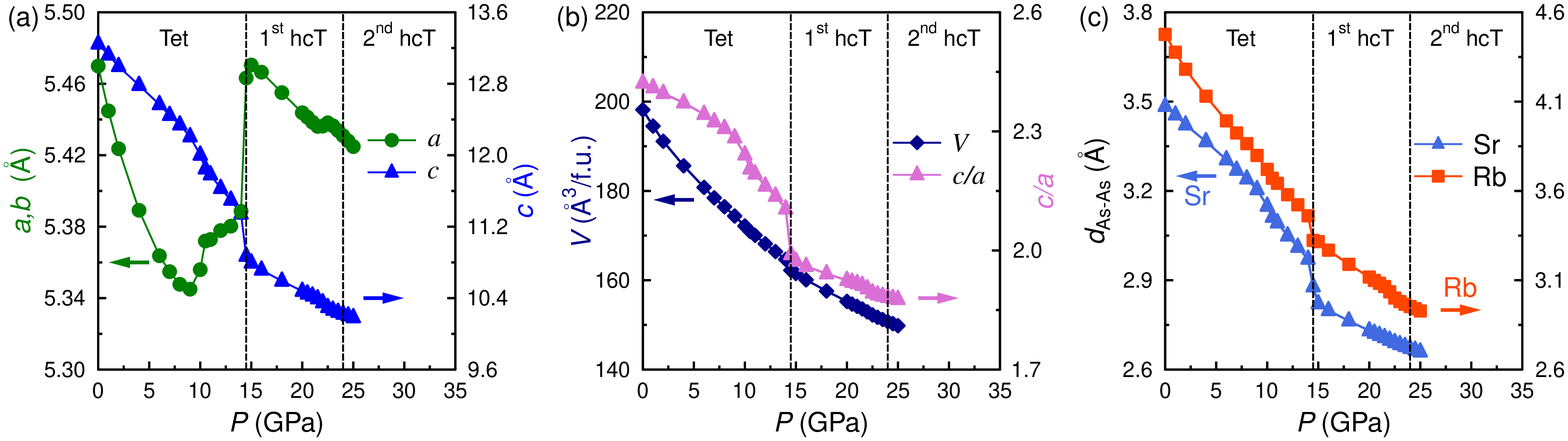}\vspace{-5pt}
  \caption{Pressure evolution of (a) lattice parameters $a=b$ and $c$, (b) volume and $c/a$ ratio and (c) As-As distances across both hcT transitions for SrRbFe$_4$As$_4$. The critical pressures of the two half-collapsed transitions are marked by vertical dashed lines. Here, the first hcT and the collapse of Fe moments occur simultaneously.} \label{f:SrRb_structure_vs_pressure}
\end{figure*}

\begin{figure*}
  \includegraphics[width = 0.99\textwidth]{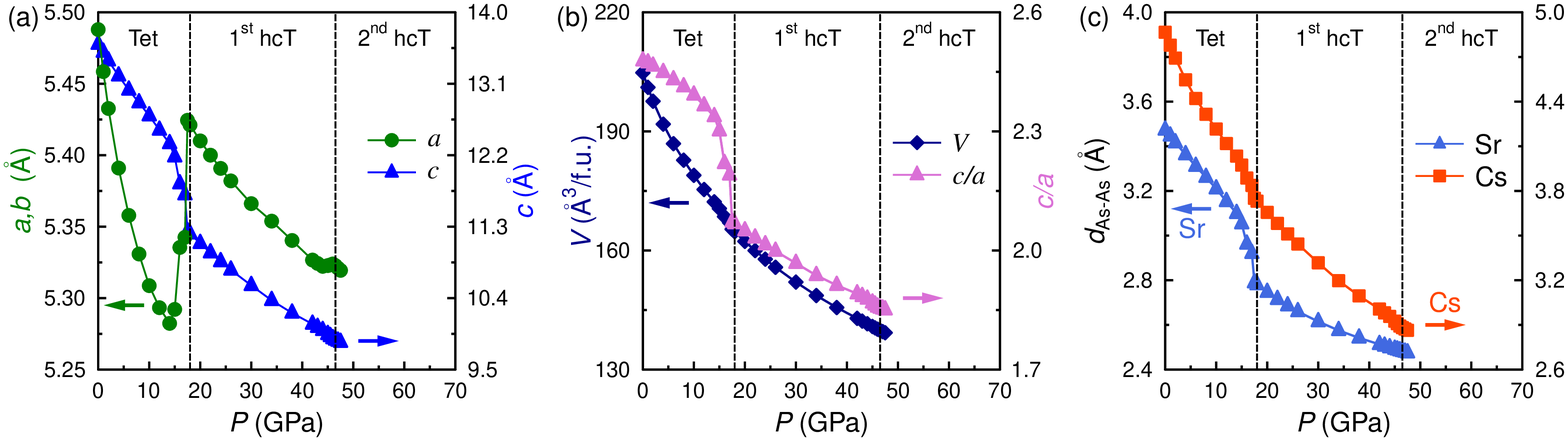}\vspace{-5pt}
  \caption{Pressure evolution of (a) lattice parameters $a=b$ and $c$, (b) volume and $c/a$ ratio and (c) As-As distances across both hcT transitions for SrCsFe$_4$As$_4$. The critical pressures of the two half-collapsed transitions are marked by vertical dashed lines. Here, the first hcT and the collapse of Fe moments occur simultaneously.} \label{f:SrCs_structure_vs_pressure}
\end{figure*}

\begin{figure*}
  \includegraphics[width = 0.99\textwidth]{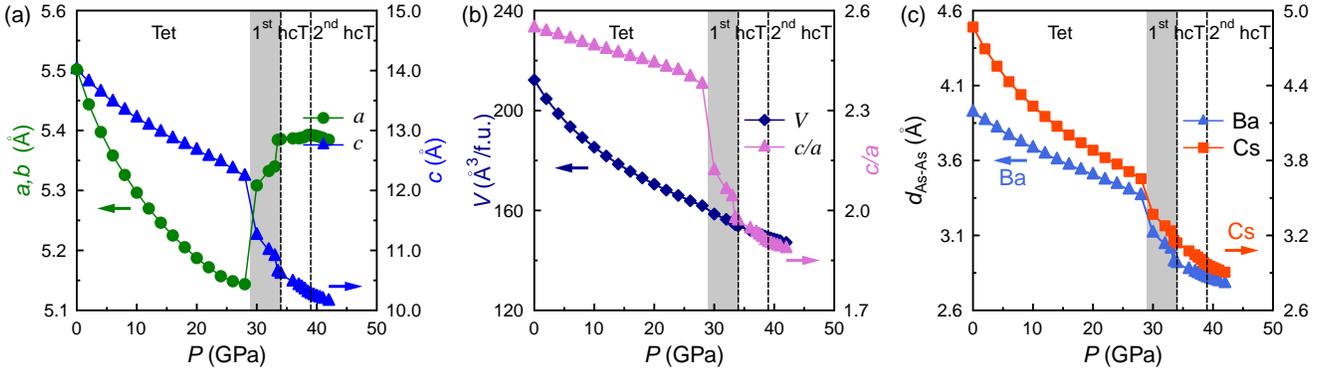}\vspace{-5pt}
  \caption{Pressure evolution of (a) lattice parameters $a=b$ and $c$, (b) volume and $c/a$ ratio and (c) As-As distances across both hcT transitions for BaCsFe$_4$As$_4$. The critical pressures of the two half-collapsed transitions are marked by vertical dashed lines and the pressure range between the collapse of Fe moments and the subsequent first hcT is indicated by shading.} \label{f:BaCs_structure_vs_pressure}
\end{figure*}

The pressure-dependence of the in-plane ($a$) and out-of-plane ($c$) lattice parameters 
along with the As-As distances across both collapse transitions is summarized for all 
studied 1144 systems in Figs.~\ref{f:CaRb_structure_vs_pressure}-\ref{f:BaCs_structure_vs_pressure}. 
At the first half collapse, the qualitative behavior is the same for all Ca\textit{A}Fe$_4$As$_4$ 
compounds, but for SrRbFe$_4$As$_4$ and SrCsFe$_4$As$_4$ the structural collapse transition overlaps 
with the suppression of Fe moments. BaCsFe$_4$As$_4$ seems to be an extreme case, since it shows 
the magnetic collapse first and then undergoes an actual half collapse at a somewhat higher pressure.

\newpage

\end{document}